\documentclass{article}

\usepackage{jheppub}

\usepackage{amsmath, amssymb, amsthm} 
\usepackage{mathtools}
\usepackage{thmtools}
\usepackage{bm}
\usepackage{dsfont}
\usepackage{braket}
\usepackage{graphicx}
\usepackage{parskip}
\usepackage{xcolor}

\numberwithin{equation}{section}
\usepackage{soul}
\usepackage{enumitem}
\setlist[itemize]{noitemsep}
\setlist[description]{noitemsep}

\usepackage[numbers, sort&compress]{natbib}
\usepackage{hyperref}
\hypersetup{
    linkcolor=[rgb]{0.91, 0.24, 0.43}, 
    citecolor=[rgb]{0.28, 0.24, 0.75}
    }

\newcommand{\dd}{\text{d}}

\newcommand{\pp}{\partial}

\newcommand{\ol}{\overline}

\newcommand{\zb}{\overline{z}}
\renewcommand{\[}{\left[}
\renewcommand{\]}{\right]}
\renewcommand{\(}{\left(}
\renewcommand{\)}{\right)}

\renewcommand{\epsilon}{\varepsilon}

\newtheorem{identity}{Identity}

\begin{document}

\date{\today}

\title{Revealing the conformal symmetry of the discrete series scalars in dS${}_2$}

\author[a]{Lukas W. Lindwasser}

\affiliation[a]{Department of Physics and Center for Theoretical Physics,
National Taiwan University, Taipei 10617, Taiwan}
    
\emailAdd{llindwasser@ntu.edu.tw}

\abstract{
On two-dimensional manifolds with nonzero constant Ricci curvature, there exists an infinite sequence of scalar fields with nonzero mass parameter that admit a pair of (anti-)holomorphic currents. After suitably defining the theory in de Sitter space ($\mathrm{dS}_2$), correlation functions of these currents obey global conformal Ward identities. We address the question of how this global conformal symmetry manifests as an action on the scalar field. An essential step is in leveraging an equivalent description of the scalar field in terms of a conformal Killing tensor. Through this, we find a conformal symmetry transformation that acts locally on the conformal Killing tensor, but non-locally on the scalar field. We show that the equation of motion transforms covariantly with respect to these conformal transformations, and further find a traceless stress tensor in both $\mathrm{dS}_2$ and $\mathrm{AdS}_2$, locally defined in terms of the conformal Killing tensor, which generates the global conformal isometry transformations.
}

\newpage

\maketitle

\section{Introduction}

Recently it was discovered that certain free massive scalar fields $\phi$ in a two-dimensional constant curvature spacetime admit a holomorphic decomposition \cite{Farnsworth:2024yeh,Chen:2026kjo}, owing to the existence of a conserved rank $k+1$ symmetric and traceless tensor
\begin{align}
    \label{eq:shiftcurrent}
    \mathcal{F}_{\mu_1\cdots\mu_{k+1}}=\nabla_{(\mu_1}\cdots\nabla_{\mu_{k+1})_T}\phi,
\end{align}
where $(\cdots)_T$ denotes the projection onto totally symmetric and traceless components. These theories are a special two-dimensional case belonging to a large class of scalar fields with conserved higher rank tensors associated with shift symmetries in (Anti-)de Sitter spacetimes \cite{Bonifacio:2018zex,Hinterbichler:2014cwa,Blauvelt:2022wwa,Bonifacio:2023prb,Hinterbichler:2024vyv}. For two-dimensional constant curvature spacetimes, there is always a choice of local isothermal complex coordinates after analytic continuation to Euclidean signature which map the surface to either the complex upper half-plane or the Riemann sphere. In each case respectively, the invariant interval takes the form
\begin{align}
    \dd s^2=\Omega(z,\zb)^2\dd z\dd\zb,\quad \Omega(z,\zb)^{2} = 
    \begin{cases}
        \frac{8}{R}\frac{1}{(z-\zb)^{2}}, & R<0\\
        \frac{8}{R}\frac{1}{(1+z\zb)^{2}}, & R>0,
    \end{cases}
    \label{eq:coords}
\end{align}
where $R$ is the Ricci scalar. In these coordinates, the conserved tensor splits up into (anti-)holomorphic currents $\mathcal{F}_{z\cdots z}\equiv F$ and $\mathcal{F}_{\zb\cdots \zb}\equiv \ol F$,
\begin{align}
\label{eq:FFb}
    F(z)=\nabla_z^{k+1}\phi,\quad \ol F(\zb)=\nabla_{\zb}^{k+1}\phi.
\end{align}
These currents become conserved when the scalar field has a special mass squared parameter proportional to the Ricci scalar,
\begin{align}
    m_k^2=-\frac{k(k+1)}{2}R,\quad k\in\mathbb{Z}_{\geq 0}.
\end{align}
These masses correspond to integer scaling dimensions $\Delta=k+1$ in $\mathrm{(A)dS}_2$. In $\mathrm{dS}_2$, the mass squared parameter is tachyonic. Despite this, the theory is unitary after defining the theory appropriately---as discussed for instance in Section \ref{ssec:dS2}---and realizes the discrete series representation of the de Sitter group, reviewed in \cite{Joung:2007je,Basile:2016aen,Sun:2021thf,Schaub:2024rnl} and also recently in \cite{Hinterbichler:2026xqf}. For this reason these scalar fields are sometimes referred to as discrete series scalar fields.

The (anti-)holomorphic components $F(z)$ and $\ol{F}(\zb)$ define chiral algebras, enhancing the local symmetry algebra of these theories beyond that of a scalar field with a more generic mass parameter. The holomorphic component $F(z)$ has modes $F_n$ which satisfy a higher derivative generalization of the $\mathrm{U}(1)$ Kac--Moody algebra,
\begin{align}
\label{eq:Fmode}
    F(z)=\frac{\mathrm{i}}{\sqrt{4\pi}}\sum_{n=-\infty}^\infty\frac{F_n}{z^{n+k+1}},\quad [F_m,F_n]=(-1)^{k}m\textstyle{\prod}_{j=1}^{k}(m^2-j^2)\delta_{m+n,0}.
\end{align}
More generally, all local holomorphic operators $\mathcal{L}(\{F(z),\pp_z F(z),\dots\},z)$ built from $F(z)$ define a symmetry of this theory. This enhancement of symmetry allows for the possibility of building local interactions which preserve integrability \cite{Zamolodchikov:1989hfa,Smirnov:2016lqw,Lindwasser:2024qyh,Lindwasser:2025slu,Chen:2026kjo}\footnote{Note that this evades the no-go theorem on integrable interacting theories in $\mathrm{AdS}_2$ \cite{Antunes:2025iaw}, which assumes that the only local higher-spin charges in a free theory are the ones present for generic mass free fields.}. 

Particularly when $k=0,1$, these theories show up in various settings. When $k=0$, the scalar field is conformally coupled and the conserved tensor \eqref{eq:shiftcurrent} $\mathcal{F}_\mu=\pp_\mu\phi$ is just the $\mathrm{U}(1)$ shift current. These scalar fields preserve the aforementioned symmetries on any Riemann surface, and are famously used to define the worldsheet action of a string embedded in flat space. When $k=1$ in $\mathrm{AdS}_2$, these fields show up as coordinate fluctuations around various minimal surface string solutions embedded in the $\mathrm{AdS}_5\times S^5$ background \cite{Drukker:2000ep,Giombi:2017cqn,Beccaria:2026ffm}. The $k=1$ fields also appear in quantum gravity coupled to conformal matter with large central charge and positive cosmological constant \cite{PhysRevD.53.3108,Anninos:2021ene}.

At least in Euclidean $\mathrm{AdS}_2$ and $\mathrm{dS}_2$, the general solution to the Klein--Gordon equation with this mass may be written in terms of purely holomorphic $W(z)$ and anti-holomorphic $\ol{W}(\zb)$ components
\begin{align}
\label{eq:holsplit1}
    \phi(z,\zb)=\mathcal{D}_k W(z) + \ol{\mathcal{D}}_k\ol{W}(\zb),
\end{align}
where $\mathcal{D}_k$ and $\ol{\mathcal{D}}_k$ are certain order-$k$ differential operators defined in Section \ref{sec: ads2}. Holomorphic splitting of the solution space is a feature normally associated with two-dimensional conformally invariant theories. Indeed, conformal invariance of the action implies a traceless stress tensor $T_{\mu\nu}$ with (anti-)holomorphic independent components $T_{zz}\equiv T(z)$ and $T_{\zb\zb}\equiv \ol T(\zb)$. If the stress tensor is a local functional of the fields, this implies there is some local operation on the fields which projects onto its (anti-)holomorphic parts.

The question of whether this theory exhibits global conformal symmetry in $\mathrm{dS}_2$ was discussed in \cite{Farnsworth:2024yeh}. In it they point out that, partly due to the (anti-)holomorphicity of the currents $F(z)$ and $\ol{F}(\zb)$, they admit two-point functions consistent with global conformal Ward identities
\begin{subequations}
\label{eq:FFtwopoint}
    \begin{align}   
    &\langle F(z) F(z')\rangle = (-1)^{k+1}\frac{1}{4\pi}(2k+1)!\frac{1}{(z-z')^{2(k+1)}},\\
    &\langle F(z) \ol F(\zb')\rangle = 0,\\
    &\langle \ol F(\zb) \ol F(\zb')\rangle = (-1)^{k+1}\frac{1}{4\pi}(2k+1)!\frac{1}{(\zb-\zb')^{2(k+1)}}.
\end{align}
\end{subequations}
How the conformal symmetry manifests as an action on the scalar field $\phi$, or if the theory admits a traceless stress tensor, was left undetermined. Clearly, the global conformal symmetry cannot act on a massive scalar field $\phi$ in the standard way, and is expected to be non-local \cite{Farnsworth:2024yeh}.

\hspace{15pt} The purpose of this paper is to present a geometric realization of the global conformal symmetry group on the scalar field $\phi$. The organization of this paper is as follows. As we describe in Section \ref{sec:scalartoCKT}, the idea is to take seriously the holomorphic splitting of the general solution \eqref{eq:holsplit1}, and interpret $W(z)$ and $\ol{W}(\zb)$ as components of a symmetric and traceless conformal Killing tensor $\mathcal{W}^{\mu_1\cdots\mu_k}$. Once this is established, there is a natural choice for how $\mathcal{W}^{\mu_1\cdots\mu_k}$ transforms under conformal symmetry transformations. In Section \ref{sec:confinv}, we show that while the action is not invariant under this choice, it leaves the solution space invariant. This is done by showing that the Klein--Gordon equation is covariant with respect to an infinitesimal conformal symmetry transformation. 

The conformal symmetry is therefore an on-shell symmetry acting on the solution space rather than an off-shell symmetry of the action. For such a symmetry, one should  not in general expect a Noether current associated with it. Despite this, in Section \ref{sec:traceless} we present an operator which plays the role of the `traceless stress tensor' $T_{\mu\nu}$ for this theory, which is a local functional of $\mathcal{W}^{\mu_1\cdots\mu_k}$. The stress tensor components $T(z)$ and $\ol{T}(\zb)$ are non-local in $\phi$, but can be used to construct well-defined charges that generate global conformal transformations. Then in Section \ref{sec: conclusion}, we conclude with comments.

In Appendix \ref{app: LtoE}, we review spacetime coordinates used to describe $\mathrm{(A)dS}_2$, and their analytic continuations to the complex coordinates \eqref{eq:coords}. The proof of the identity \eqref{eq:KGtoCKT}, which is crucial for the equivalence between the discrete series scalar field and a conformal Killing tensor, is presented in Appendix \ref{app: KGtoCKT}. Finally in Appendix \ref{app:b=0}, we provide further technical details on the monodromy relations of the stress tensor components in $\mathrm{dS}_2$.

\subsection*{Conventions}
Although we present covariant formulae in Lorentzian signature, when using an explicit coordinate system we for the most part be in Euclidean signature, with the complex coordinates \eqref{eq:coords}, depending on whether the Ricci scalar is negative or positive. Complex partial derivatives $\pp_z$ ($\pp_{\zb}$) are denoted by $\pp$ ($\ol{\pp}$). We also refer to both Lorentzian and Euclidean signature (anti-)de Sitter spaces as $\mathrm{(A)dS}_2$. Tensor indices are symmetrized with unit weight, e.g. $\mathcal{O}_{(\mu\nu)}\equiv\frac{1}{2}(\mathcal{O}_{\mu\nu}+\mathcal{O}_{\nu\mu})$, and $(\cdots)_T$ denotes traceless symmetrization, e.g. $\mathcal{O}_{(\mu\nu)_T}\equiv\frac{1}{2}(\mathcal{O}_{\mu\nu}+\mathcal{O}_{\nu\mu})-\frac{1}{2}g_{\mu\nu}\mathcal{O}^{\lambda}_{\;\;\lambda}$.


\section{The scalar field as a conformal Killing tensor \label{sec: ads2}}
\label{sec:scalartoCKT}
In \cite{Chen:2026kjo}, it was shown that the general solution to the Klein--Gordon equation with mass squared $m_k^2=-k(k+1)R/2$, assuming either vanishing Dirichlet conformal boundary conditions in the case of $\mathrm{AdS}_2$, or periodic boundary conditions in $\mathrm{dS}_2$, can be expressed in the complex coordinates \eqref{eq:coords} as
\begin{align}
\label{eq:noncovAnsatz}
    \phi(z,\zb)=\mathcal{D}_kW(z) + \ol{\mathcal{D}}_k\ol{W}(\zb).
\end{align}
where $\mathcal{D}_k$ and $\ol{\mathcal{D}}_k$ are order-$k$ differential operators equal to
\begin{align}
        &\mathcal{D}_{k} \cdot \equiv  \(\pp + 2\frac{\pp\Omega}{\Omega}\)\(\pp + 4\frac{\pp\Omega}{\Omega}\) \dots \(\pp + 2k\frac{\pp\Omega}{\Omega}\)\cdot,\;\ol{\mathcal{D}}_{k} \cdot \equiv  \(\ol\pp + 2\frac{\ol\pp\Omega}{\Omega}\)\(\ol\pp + 4\frac{\ol\pp\Omega}{\Omega}\) \dots \(\ol\pp + 2k\frac{\ol\pp\Omega}{\Omega}\)\cdot
\end{align}
A crucial observation\footnote{This observation was originally made as a footnote in \cite{Chen:2026kjo}.} for this paper is that these differential operators are precisely those that appear in the $k$-fold divergence of a rank $k$ symmetric and traceless contravariant tensor $\mathcal{W}^{\mu_1\cdots\mu_k}$, i.e.
\begin{align}
    \label{eq:divergenceAnsatz}
    \phi(z,\zb)=\nabla_{\mu_1}\cdots\nabla_{\mu_k}\mathcal{W}^{\mu_1\cdots\mu_k}.
\end{align}
where the independent components of $\mathcal{W}^{\mu_1\cdots\mu_k}$ are
\begin{align}
    \mathcal{W}^{z\cdots z}\equiv W(z),\quad \mathcal{W}^{\zb\cdots\zb}\equiv\ol{W}(\zb).
\end{align}
A symmetric and traceless contravariant tensor with (anti-)holomorphic components such as this solves the conformal Killing tensor equation
\begin{align}
\label{eq:CKT}
    \nabla_{(\mu_1}\mathcal{W}_{\mu_2\cdots\mu_{k+1})_T}=0.
\end{align}
It is instructive to see how to recover the conformal Killing tensor equation from the Klein--Gordon equation. Using the Ansatz \eqref{eq:divergenceAnsatz}, we show in Appendix \ref{app: KGtoCKT} that the following intertwining relation holds for two-dimensional constant curvature spacetimes and symmetric and traceless tensors $\mathcal{W}^{\mu_1\cdots\mu_k}$
\begin{align}
\label{eq:KGtoCKT}
    \(\Delta+\frac{k(k+1)}{2}R\)\nabla_{\mu_1}\cdots\nabla_{\mu_k}\mathcal{W}^{\mu_1\cdots\mu_k}=2\nabla_{\mu_1}\cdots\nabla_{\mu_{k+1}}\nabla^{(\mu_1}\mathcal{W}^{\mu_2\cdots\mu_{k+1})_T}.
\end{align}
This relation guarantees that $\phi$ as obtained from a conformal Killing tensor solves the Klein--Gordon equation. There are clearly more local solutions to the Klein--Gordon equation than this. Indeed, the tensor $\nabla^{(\mu_1}\mathcal{W}^{\mu_2\cdots\mu_{k+1})_T}$ could instead be divergenceless, or most generally have zero $(k+1)$-fold divergence, with $\phi$ still solving the Klein--Gordon equation. After imposing appropriate boundary conditions\footnote{ In the case of $\mathrm{AdS}_2$ we must impose vanishing Dirichlet conformal boundary conditions when $k>0$ \cite{Higuchi:2021fxg}. In $\mathrm{dS}_2$, we impose periodic boundary conditions. If $\phi$ is complex, there are more general boundary conditions in $\mathrm{dS}_2$ which we do not consider \cite{Higuchi:2022nfy}.} however, every solution can be written in terms of $\mathcal{W}^{\mu_1\cdots\mu_k}$ as a conformal Killing tensor \cite{Chen:2026kjo}. This means that $\phi$ has an equivalent description in terms of the conformal Killing tensor $\mathcal{W}^{\mu_1\cdots\mu_k}$.

Using \eqref{eq:KGtoCKT}, we rewrite the action manifestly in terms of $\mathcal{F}_{\mu_1\cdots\mu_{k+1}}$ and $\mathcal{W}^{\mu_1\cdots\mu_k}$ after repeated integration by parts, up to boundary terms,
\begin{subequations}
\begin{align}
    S&=-\frac{1}{2}\int \dd^2x\sqrt{-g}\left(g^{\mu\nu}\pp_\mu\phi\pp_\nu\phi-\frac{k(k+1)}{2}R\phi^2\right)\\
    &=(-1)^{k+1}\int \dd^2x\sqrt{-g}\mathcal{F}_{\mu_1\cdots\mu_{k+1}}\nabla^{(\mu_1}\mathcal{W}^{\mu_2\cdots\mu_{k+1})_T} +\dots
    \label{eq:FWaction}
\end{align}
\end{subequations}
\section{Conformal invariance}
\label{sec:confinv}
The conformal Killing tensor equation \eqref{eq:CKT} has special properties under Weyl rescaling. If a spacetime metric $g$ admits a conformal Killing tensor $\mathcal{W}_{\mu_1\cdots\mu_k}$, then a conformally related spacetime $\hat g=e^{2\sigma}g$ admits the conformal Killing tensor $\hat{\mathcal{W}}_{\mu_1\cdots\mu_k}=e^{2k\sigma}\mathcal{W}_{\mu_1\cdots\mu_k}$. In terms of the contravariant tensor, this is the statement that $\hat{\mathcal{W}}^{\mu_1\cdots\mu_k}=\mathcal{W}^{\mu_1\cdots\mu_k}$ is Weyl invariant. This suggests a natural classical action of conformal symmetry transformations on $\phi$ in terms of the conformal properties of $\mathcal{W}^{\mu_1\cdots\mu_k}$.

\subsection{Conformal isometries in $\mathrm{(A)dS}_2$}

A conformal symmetry transformation is a simultaneous conformal coordinate transformation via a Lie derivative $\mathcal{L}_\xi$, and Weyl rescaling such that the metric is left invariant,
\begin{align}
    (\delta_\xi+\delta_\sigma)g_{\mu\nu}&=-\mathcal{L}_\xi g_{\mu\nu}+2\sigma g_{\mu\nu}\\
    \label{eq:CKV}
    &=-\nabla_\mu\xi_\nu-\nabla_\nu\xi_\mu + 2\sigma g_{\mu\nu}=0
\end{align}
so that  $\sigma=\nabla_\mu\xi^\mu/2$. $\xi^{\mu}$ must therefore solve the conformal Killing vector equation, which defines the conformal isometries of the spacetime. In two dimensions, there are infinitely many local solutions $\xi^\mu$ to \eqref{eq:CKV}, which in the complex coordinates \eqref{eq:coords} are parameterized by (anti-)holomorphic components $\xi^z=\xi(z)$, and $\xi^{\zb}=\ol\xi(\zb)$. 

For globally well-defined coordinate transformations, $\xi(z)$ and $\ol\xi(\zb)$ are restricted to be quadratic functions of $z$ and $\zb$, respectively
\begin{align}
    \xi(z)=a+bz+cz^2,\quad\ol\xi(\zb)=\ol a+\ol b\zb+\ol c\zb^2.
    \label{eq:globalCKV}
\end{align}
In $\mathrm{AdS}_2$, the conformal Killing vector must preserve the upper half-plane, which enforces $a,b,c\in\mathbb{R}$. These are precisely the isometries of the upper half-plane, meaning that $\sigma=0$. In this sense, any quantum field theory in $\mathrm{AdS}_2$ is invariant under global conformal isometries.

In $\mathrm{dS}_2$, there are more global conformal isometries than isometries. Indeed, the isometries of $\mathrm{dS}_2$ take the form \eqref{eq:globalCKV} with $a=\ol c$ and $b=-\ol b$, while global conformal isometries allow for $a,b,c$ to be independent complex parameters. In $\mathrm{dS}_2$, the global conformal Weyl parameter $\sigma$ is explicitly,
\begin{align}
    \sigma_{\mathrm{dS}}=\frac{1}{2(1+z\zb)}\(\(b+\ol b\)\(1-z\zb\)+2z\(c-\ol{a}\)+2\zb\(\ol{c}-a\)\).
\end{align}

A covariant identity we will use in the following that global conformal isometries obey is a condition on $\sigma$
\begin{align}
\label{eq:GlobalConf}
    \nabla_{(\mu}\nabla_{\nu)_T}\sigma=0,
\end{align}
which in the complex coordinates \eqref{eq:coords} means that $\nabla_z^2\sigma=\nabla_{\zb}^2\sigma=0$.

The distinction we make above and throughout between global and local conformal isometries is particular to the Euclidean signature spaces. In their corresponding Lorentzian signature spacetimes, a different set of conformal isometries may be globally well-defined. For instance, in $\mathrm{dS}_2$ a much larger set of conformal isometries $\xi^\mu(\eta,\vartheta)$ in the coordinates defined in Appendix \ref{sapp:dS} are globally well-defined, with $\xi^\eta=f(\vartheta+\eta)-f(\vartheta-\eta+\pi)$ and $\xi^{\vartheta}=f(\vartheta+\eta)+f(\vartheta-\eta+\pi)$ with $f(x+2\pi)=f(x)$ periodic. It turns out however that the criterion we use for global conformal isometries in Euclidean signature naturally picks out the group of conformal isometries that leaves the vacuum defined in \eqref{eq:vacuum} invariant in $\mathrm{dS}_2$.
\subsection{Conformal symmetry transformation}
\label{ssec:cst}
If we suppose $\mathcal{W}^{\mu_1\cdots\mu_k}$ is Weyl invariant, then the conformal symmetry transformation with respect to an arbitrary local solution to \eqref{eq:CKV} of $\phi$ is,
\begin{align}
    (\delta_\xi+\delta_\sigma)\phi&=-\mathcal{L}_\xi\phi+\sum_{i=0}^k2i{k+1\choose i+1}\nabla_{\mu_1}\cdots\nabla_{\mu_i}\sigma \nabla_{\mu_{i+1}}\cdots\nabla_{\mu_k}\mathcal{W}^{\mu_1\cdots\mu_k}
    \label{eq:phiconftran}
\end{align}
For global conformal isometries \eqref{eq:GlobalConf}, only the $i=1$ term within the sum is nonzero, due to the fact that $\mathcal{W}^{\mu_1\cdots\mu_k}$ is symmetric and traceless,
\begin{align}
    (\delta_\xi+\delta_{\sigma})\phi=-\mathcal{L}_\xi\phi+k(k+1)\nabla_{\mu_1}\sigma\nabla_{\mu_2}\cdots\nabla_{\mu_k}\mathcal{W}^{\mu_1\cdots\mu_k},\quad \text{(Global conformal isometry)}.
    \label{eq:phiglobconftran}
\end{align}
Note that this transformation acts on $\phi$ in the standard way in $\mathrm{AdS}_2$ because $\sigma=0$, and is strictly non-local as a functional of $\phi$ in $\mathrm{dS}_2$ (as long as $\sigma\neq 0$), defined instead locally with respect to $\mathcal{W}^{\mu_1\cdots\mu_k}$. 

After the equation of motion is imposed, the components $\mathcal{F}_{z\cdots z}=F(z)$ and $\mathcal{F}_{\zb\cdots\zb}=\ol F(\zb)$ \eqref{eq:FFb} are (anti-)holomorphic, and have the following expression in terms of $W(z)$ and $\ol{W}(\zb)$, respectively \cite{Chen:2026kjo}, 
\begin{align}
\label{eq:FtoWcov}
    F(z)=\nabla_z^{2k+1}W(z),\quad \ol{F}(\zb)=\nabla_{\zb}^{2k+1}\ol{W}(\zb).
\end{align}
That these are (anti-)holomorphic on a constant curvature background is a non-trivial geometric statement. Indeed, using either coordinates in \eqref{eq:coords}, the covariant derivatives in \eqref{eq:FtoWcov} can be replaced with partial derivatives \cite{Chen:2026kjo},
\begin{align}
    F(z)=\pp^{2k+1}W(z),\quad \ol{F}(\zb)=\ol\pp^{2k+1}\ol{W}(\zb).
\end{align}
This formula suggests that $F(z)$ and $\ol{F}(\zb)$ are themselves Weyl invariant, given that all metric dependence from covariant differentiation vanishes\footnote{In a more general complex coordinate system, the operators $\pp^{2k+1}$ and $\ol{\pp}^{2k+1}$ should be replaced with the corresponding order $2k+1$ Bol operators \cite{Bol1949,doi:10.1142/S0217751X93000023}, which have Weyl rescaling dependence through the `projective connection'.}. This is however not true for a general Weyl transformation. Using \eqref{eq:FtoWcov}, we find 
\begin{align}
\label{eq:FFbWeyl}
    \delta_\sigma F=\sum_{i=2}^{2k+1}(i-1){2k+2\choose i+1}\nabla_z^i\sigma\nabla_z^{2k+1-i}W(z),\quad \delta_\sigma \ol F=\sum_{i=2}^{2k+1}(i-1){2k+2\choose i+1}\nabla_{\zb}^i\sigma\nabla_{\zb}^{2k+1-i}\ol{W}(\zb).
\end{align}
These Weyl variations depend only on $\nabla_z^i\sigma$ and $\nabla_{\zb}^i\sigma$ for $i\geq 2$, which all vanish for global Weyl transformations \eqref{eq:GlobalConf}. Because of this, $\mathcal{F}_{\mu_1\cdots\mu_{k+1}}$ is globally Weyl invariant on a constant curvature background, after the equation of motion is imposed. In this sense, $\mathcal{F}_{\mu_1\cdots\mu_{k+1}}$ is a primary with respect to global conformal transformations. In terms of its components $F(z)$ and $\ol{F}(\zb)$, they are global primaries with holomorphic weights $(h_F,\ol{h}_{F})=(k+1,0)$ and $(h_{\ol F},\ol{h}_{\ol{F}})=(0,k+1)$, respectively. If furthermore the vacuum state is invariant under the global conformal transformations \eqref{eq:GlobalConf}, then the transformation \eqref{eq:FFbWeyl} implies the Ward identities that fix the two-point functions for $F(z)$ and $\ol{F}(\zb)$ to be of the form \eqref{eq:FFtwopoint}. We will show that the vacuum state is invariant under these transformations in Section \ref{ssec:dS2}.

Because $\mathcal{F}_{\mu_1\cdots\mu_{k+1}}$ is only a primary after imposing the equation of motion, the action \eqref{eq:FWaction} as written in terms of $\mathcal{F}_{\mu_1\cdots\mu_{k+1}}$ and $\mathcal{W}^{\mu_1\cdots\mu_k}$ is neither invariant under global nor local conformal transformations. Instead, the conformal symmetry is manifest on the space of solutions, where the boundary conditions constrain $\mathcal{W}^{\mu_1\cdots\mu_k}$ to be a conformal Killing tensor. The equation of motion as obtained from varying the action with respect to $\phi$ is covariant with respect to arbitrary conformal transformations \eqref{eq:phiconftran}. Indeed, once we write the Klein--Gordon equation in terms of the conformal Killing tensor equation as in the right hand side of \eqref{eq:KGtoCKT}, it is clear that the transformation \eqref{eq:phiconftran} preserves the solution space, due to the conformal Killing tensor equation itself having definite Weyl weight $-2$. Explicitly, the equation of motion transforms infinitesimally as
\begin{align}
    &(\delta_\xi+\delta_\sigma)\(\nabla_{\mu_1}\cdots\nabla_{\mu_{k+1}}\mathcal{K}^{\mu_1\cdots\mu_{k+1}}\)=\nonumber\\
    &-\mathcal{L}_\xi\(\nabla_{\mu_1}\cdots\nabla_{\mu_{k+1}}\mathcal{K}^{\mu_1\cdots\mu_{k+1}}\)+ 
    \sum_{i=0}^{k+1}2\frac{i(k+1)-1}{i+1}{k+1\choose i}\nabla_{\mu_1}\cdots\nabla_{\mu_i}\sigma \nabla_{\mu_{i+1}}\cdots\nabla_{\mu_{k+1}}\mathcal{K}^{\mu_1\cdots\mu_{k+1}},
\end{align}
where $\mathcal{K}^{\mu_1\cdots\mu_{k+1}}=\nabla^{(\mu_1}\mathcal{W}^{\mu_2\cdots\mu_{k+1})_T}$. This variation vanishes because the equation of motion together with the appropriate boundary condition implies $\mathcal{K}^{\mu_1\cdots\mu_{k+1}}=0$.

\section{A traceless stress tensor}
\label{sec:traceless}
The standard symmetric stress tensor $T_{\mu\nu}$ is defined by varying the action with respect to the inverse metric $g^{\mu\nu}$, assuming $\phi$ has no metric dependence. Instead, we assume $\mathcal{W}^{\mu_1\cdots\mu_k}$ has no metric dependence and vary \eqref{eq:FWaction} with respect to the inverse metric $g^{\mu\nu}$,
\begin{align}
\label{eq:Tdefn}
    \delta_g S=-\frac{1}{2}\int \mathrm{d}^2x\sqrt{-g}\,\delta g^{\mu\nu}T_{\mu\nu}.
\end{align}
An infinitesimal Weyl variation depends only on the trace $T^{\mu}_{\;\;\mu}$
\begin{align}
    \delta_\sigma S=\int \mathrm{d}^2x\sqrt{-g}\,\sigma T^{\mu}_{\;\;\mu}.
\end{align}
In constant curvature spacetimes, due to the global Weyl parameter satisfying \eqref{eq:GlobalConf}, the action is globally Weyl invariant as long as $T^{\mu}_{\;\;\mu}=\nabla_{(\mu}\nabla_{\nu)} L^{\mu\nu}+\frac{1}{2}RL^{\mu}_{\;\;\mu}$, with $L^{\mu\nu}$ some contravariant rank $2$ tensor. For the action to be invariant under arbitrary local Weyl transformations, the trace can also be nonzero $T^{\mu}_{\;\;\mu}=(\Delta+R)L$, where $L$ is some scalar function. For the latter case, it is always possible to find an improvement term which makes $T_{\mu\nu}'=T_{\mu\nu}+\Delta T_{\mu\nu}$ identically traceless,
\begin{align}
    T_{\mu\nu}'= T_{\mu\nu}+(\nabla_\mu\nabla_\nu-g_{\mu\nu}\Delta)L-\frac{1}{2}Rg_{\mu\nu}L.
\end{align}
As discussed in Section \ref{ssec:cst}, because $\mathcal{F}_{\mu_1\cdots\mu_{k+1}}$ behaves like a primary under the global conformal transformation \eqref{eq:phiglobconftran} only after the equation of motion is imposed, the action is neither globally nor locally Weyl invariant in the sense we proposed. One therefore should not expect to find a stress tensor that is identically traceless.

Despite this, we find that the stress tensor as defined in \eqref{eq:Tdefn} is traceless after the equation of motion is imposed. This stress tensor is local in $\mathcal{W}^{\mu_1\cdots\mu_k}$ (i.e. non-local in $\phi$), and we show that it generates global conformal symmetry transformations in Sections \ref{sssec:AdS2} and \ref{ssec:dS2}. For $k>0$, we obtain this stress tensor by taking the functional derivative of \eqref{eq:FWaction} with respect to $g^{\mu\nu}$, ignoring terms which vanish due to the equation of motion as well as boundary terms,
\begin{align}
\label{eq:Tmunu}
    T_{\mu\nu}=(-1)^{k}\(2\mathcal{F}_{\lambda_1\cdots\lambda_k(\mu}\nabla_{\nu)}\mathcal{W}^{\lambda_1\cdots\lambda_k}+(k-1)\mathcal{F}_{\mu\nu\lambda_1\cdots\lambda_{k-1}}\nabla_{\lambda_k}\mathcal{W}^{\lambda_1\cdots\lambda_k}+k\nabla_{\lambda_{k}}\mathcal{F}_{\mu\nu\lambda_1\cdots\lambda_{k-1}}\mathcal{W}^{\lambda_1\cdots\lambda_k}\).
\end{align}
We first show that this operator is conserved and traceless after the equation of motion is imposed. To do this, we pass over to complex coordinates, noting that the components $T_{zz}$, $T_{\zb\zb}$ are (anti-)holomorphic respectively,
\begin{align}
\label{eq:Tzzhol}
    &T_{zz}=(-1)^k\[(k+1)F\pp W+k(\pp F)W\],\quad T_{\zb\zb}=(-1)^k\[(k+1)\ol F\ol\pp\ol{W}+k(\ol\pp\ol F)\ol{W}\],
\end{align}
where the Levi-Civita connections cancel between the two terms because of the tensor origins of $F,W$ and $\ol F,\ol{W}$. Furthermore, $T_{z\zb}=0$ because $\mathcal{W}^{\mu_1\cdots\mu_k}$ is a conformal Killing tensor, 
\begin{align}
\label{eq:Tzzb0}
    T_{z\zb}=(-1)^k\(F\ol\pp W+\ol F\pp \ol{W}\)=0.
\end{align}
The equations \eqref{eq:Tzzhol}--\eqref{eq:Tzzb0} together show that $T_{\mu\nu}$ is conserved and traceless. Note that the conservation and tracelessness depends not only on the equation of motion, but also the boundary condition which together forces $\mathcal{W}^{\mu_1\cdots\mu_k}$ to be a conformal Killing tensor. In the following we will denote the nonzero independent components of the stress tensor as $T_{zz}\equiv T(z)$ and $T_{\zb\zb}\equiv\ol T(\zb)$.
\subsection{Conformal generators}
In this Section, we compute the global charges associated with the currents $z^{m+1}T(z)$ and $\zb^{m+1}\ol{T}(\zb)$, and show that they reproduce the charges for conformal isometries when $m=-1,0,1$. When $|m|>1$, the charges \textit{do not} correspond to Virasoro generators, and instead obey a more complicated algebra which we omit. We present the analysis for both $\mathrm{AdS}_2$ and $\mathrm{dS}_2$, even though for $\mathrm{AdS}_2$ there is no further enhancement of global conformal symmetry.
\subsubsection{Anti-de Sitter}
\label{sssec:AdS2}
Working in the complex upper half-plane, the relevant mode expansions for building the conformal generators from \eqref{eq:Tzzhol} in $\mathrm{AdS}_2$ are \cite{Chen:2026kjo},
\begin{subequations}
\begin{align}
    W(z) =&-\frac{(-\mathrm{i})^{k+1}}{\sqrt{4\pi}}\sum_{|n|>k}\frac{1}{n}\frac{1}{\sqrt{\prod_{j=1}^k(n^2-j^2)}}\frac{\alpha_n}{z^{n-k}}, \\
        \ol{W}(\zb) =&-\frac{\mathrm{i}^{k+1}}{\sqrt{4\pi}}\sum_{|n|>k}\frac{1}{n}\frac{1}{\sqrt{\prod_{j=1}^k(n^2-j^2)}}\frac{\alpha_n}{\zb^{n-k}},\\
    F(z)=&\,\frac{(-\mathrm{i})^{k+1}}{\sqrt{4\pi}}\sum_{|n|>k} \sqrt{\textstyle\prod_{j=1}^k(n^2-j^2)} \frac{\alpha_{n}}{z^{n+k+1}},\\
        \ol F(\zb)=&\,\frac{\mathrm{i}^{k+1}}{\sqrt{4\pi}}\sum_{|n|>k} \sqrt{\textstyle\prod_{j=1}^k(n^2-j^2)} \frac{\alpha_{n}}{\zb^{n+k+1}},
\end{align}
\end{subequations}
written in terms of oscillator operators $\alpha_n$ for $|n|>k$, where the label $n$ denotes the definite energy of the corresponding mode. These are conventionally normalized with respect to the canonical creation and annihilation operators $\alpha_n=\sqrt{n}a_n$, and $\alpha_{-n}=\sqrt{n}a_n^\dagger$ with $n>k$ such that they satisfy the commutation relations $[\alpha_m,\alpha_n]=m\delta_{m+n,0}$. In terms of these operators, the stress tensor has the following Laurent expansion
\begin{align}
    T(z)=-\frac{1}{2\pi}\sum_{m=-\infty}^{\infty}\frac{\ell_m}{z^{m+2}},\quad \ol{T}(\zb)=-\frac{1}{2\pi}\sum_{m=-\infty}^{\infty}\frac{\ell_m}{\zb^{m+2}},
\end{align}
where
\begin{align}
    \ell_m=\sum_{|n|>k}\frac{n+km}{2n}\sqrt{\prod_{j=1}^k\(\frac{(m-n)^2-j^2}{n^2-j^2}\)}:\alpha_{m-n}\alpha_n:,
\end{align}
and it is understood that $\alpha_m\equiv 0$ whenever $|m|\leq k$ in this sum. These generators match exactly the generators of the isometries of $\mathrm{AdS}_2$ when $m=-1,0,1$, up to the normal ordering prescription $a_{\mathrm{AdS}}$,
\begin{subequations}
    \begin{align}
        \ell_{-1}&=\sum_{n=k+1}^\infty\sqrt{\frac{(n-k)(n+k+1)}{n(n+1)}}\alpha_{-1-n}\alpha_n,\\
        \ell_0&=\sum_{n=k+1}^\infty\alpha_{-n}\alpha_n+a_{\mathrm{AdS}},\\
        \ell_{+1}&=\sum_{n=k+1}^\infty\sqrt{\frac{(n+k)(n-k-1)}{n(n-1)}}\alpha_{1-n}\alpha_{n}.
    \end{align}
    \label{eq: AdS ell op}%
\end{subequations}
The isometries satisfy the Lie algebra of $\mathrm{SL}(2,\mathbb{R})$
\begin{align}
    [\ell_m,\ell_n]=(m-n)\(\ell_{m+n}-a_{\mathrm{AdS}}\delta_{m+n,0}\),\quad m,n=-1,0,1.
\end{align}
That the stress tensor \eqref{eq:Tmunu} correctly reproduces the isometries could have been anticipated from the fact that it is the Noether current associated with the transformation \eqref{eq:phiglobconftran}, which applies to ordinary isometries when $\sigma=0$, in which case \eqref{eq:phiglobconftran} is the usual way a scalar field $\phi$ transforms.

When $|m|>1$, the $\ell_m$ generators do not obey a closed algebra, and do not implement the local conformal transformation \eqref{eq:phiconftran}. This indicates that although formally \eqref{eq:phiconftran} is a symmetry of the equation of motion, there are no operators acting on the Hilbert space that implement it. In fact, there is no set of operators $L_m$ with $|m|>1$ that can simultaneously implement the expected transformation on $W(z)$ as a conformal primary with (anti-)holomorphic weights $(h_W,\ol{h}_W)=(-k,0)$,
\begin{align}
    [L_m,W(z)]=z^{m+1}\pp W(z)-k(m+1)z^mW(z),\quad |m|>1,
\end{align}
and the condition that $L_m$ preserves the canonical commutation relations, i.e. $[L_m,[\alpha_s,\alpha_t]]=0$ for any $s,t$ and $|m|>1$, consistent with the Jacobi identity. This statement applies equally to the theory in $\mathrm{dS}_2$, as we will describe in the next Section. In Section \ref{sec: conclusion}, we discuss a way to modify the theory to allow for the full Virasoro algebra to be realized.
\subsubsection{de Sitter}
\label{ssec:dS2}
In $\mathrm{dS}_2$ there are several features which make the stress tensor, and the theory itself, more subtle to define. We begin this Section by reviewing these subtleties, discussed in e.g. \cite{Kirsten:1993ug,Tolley:2001gg,Anninos:2023lin,Chen:2026kjo}. Working on the Riemann sphere, the relevant mode expansions for building the conformal generators in $\mathrm{dS}_2$ are 
\begin{subequations}
    \begin{align}
       W(z) =&\frac{1}{\sqrt{4\pi}}\sum_{|n|\leq k}\(\frac{1}{2}\mathrm{i}^{k+n}\frac{x_n}{z^{n-k}}+\frac{(-\mathrm{i})^{k+1+n}}
        {(k-n)!(k+n)!}\frac{p_n}{z^{n-k}}\log z
        \)+\frac{\mathrm{i}^{k+1}}{\sqrt{4\pi}}\sum_{|n|>k}\frac{1}{n}\frac{1}{\sqrt{\textstyle\prod_{j=1}^k(n^2-j^2)}}\frac{\alpha_n}{z^{n-k}}, \\
        \ol{W}(\zb) =&\frac{1}{\sqrt{4\pi}}\sum_{|n|\leq k}\(\frac{1}{2}\mathrm{i}^{k+n}\frac{x_{-n}}{\ol z^{n-k}}+\frac{(-\mathrm{i})^{k+1+n}}
        {(k-n)!(k+n)!}\frac{p_{-n}}{\ol z^{n-k}}\log \ol z\)+\frac{\mathrm{i}^{k+1}}{\sqrt{4\pi}}\sum_{|n|>k}\frac{1}{n}\frac{1}{\sqrt{\textstyle\prod_{j=1}^k(n^2-j^2)}}\frac{\tilde{\alpha}_n}{\ol z^{n-k}},\\
        F(z)=&-\frac{1}{\sqrt{4\pi}}\sum_{|n|\leq k}\mathrm{i}^{k+1+n}\frac{p_n}{z^{n+k+1}}-\frac{\mathrm{i}^{k+1}}{\sqrt{4\pi}}\sum_{|n|>k}\sqrt{\textstyle\prod_{j=1}^k(n^2-j^2)} \frac{\alpha_{n}}{z^{n+k+1}},\\
        \ol F(\ol z)=&-\frac{1}{\sqrt{4\pi}}\sum_{|n|\leq k}\mathrm{i}^{k+1+n}\frac{ p_{-n}}{\ol z^{n+k+1}}-\frac{\mathrm{i}^{k+1}}{\sqrt{4\pi}}\sum_{|n|>k} \sqrt{\textstyle\prod_{j=1}^k(n^2-j^2)} \frac{\tilde{\alpha}_{n}}{\ol z^{n+k+1}}.
    \end{align}
\end{subequations}
In this case there are two independent sets of oscillator operators $\alpha_n,\tilde{\alpha}_n$ for $|n|>k$, where the label $n$ denotes the definite angular momentum of the corresponding mode. These operators are similarly normalized such that they satisfy the commutation relations $[\alpha_m,\alpha_n]=m\delta_{m+n,0}$, $[\tilde{\alpha}_m,\tilde{\alpha}_n]=m\delta_{m+n,0}$, and $[\alpha_m,\tilde{\alpha}_n]=0$. 

There are also $x_n$ and $p_n$ operators associated with modes with $|n|\leq k$ satisfying the commutation relations $[x_m,p_n]=\mathrm{i}\delta_{m+n,0}$, $[x_m,x_n]=0$, and $[p_m,p_n]=0$. These modes have zero norm with respect to the Klein--Gordon inner product \eqref{eq:dSinnerproduct}. The $x_n$ operators are associated with globally defined solutions of the Klein--Gordon equation. The $p_n$ operators therefore implement shifts of the scalar field $\phi$ by these globally defined solutions. The existence of globally defined solutions means that the Klein--Gordon operator in $\mathrm{dS}_2$ has zero eigenvalues, which with mass squared $m_k^2=-k(k+1)R/2$ has a degeneracy $2k+1$. This obstructs a consistent quantization of the theory, because there are modes which have zero contribution to the action, making the partition function ill-defined.

The most straightforward way to make this theory well-defined is by considering how to define the $\mathrm{dS}_2$ invariant vacuum state and its associated Fock space. Without removing either the $x_n$ or $p_n$ operators from the operator algebra, any choice of definition of the vacuum state $|\Omega\rangle$ will change after the application of a $\mathrm{dS}_2$ isometry. To define a $\mathrm{dS}_2$ invariant theory then, it is necessary to remove these operators in some consistent way. The simplest way to do this is to restrict the operator algebra to consist only of operators that are invariant under shifts by the $x_n$ modes\footnote{More generally we may impose the weaker condition that the transformation under shifts $\delta_n\mathcal{O}\equiv \mathrm{i}[p_n,\mathcal{O}]$ has zero transition amplitude between physical states.}, such as $F(z)$ and $\ol{F}(\zb)$, and choose a vacuum state which is annihilated by all $p_n$, along with the oscillator operators $\alpha_n,\tilde{\alpha}_n$ for all $n>k$,
\begin{subequations}
\label{eq:vacuum}
\begin{align}
    &p_n|\Omega\rangle=0,\quad |n|\leq k,\\
    &\alpha_n|\Omega\rangle=0,\quad n>k,\\
    &\tilde\alpha_n|\Omega\rangle=0,\quad n>k.
\end{align}
\end{subequations}
All physical states in the Hilbert space are then obtained by successive application of $\alpha_{-n}$ and $\tilde{\alpha}_{-n}$ with $n>k$. Thus, physical states satisfy $p_n|\text{phys}\rangle=0$ for all $|n|\leq k$. Once we restrict to these operators and states, $p_n=0$ for all $|n|\leq k$ as operators acting on this Hilbert space. 

The components $W(z)$ and $\ol{W}(\zb)$ are neither shift invariant nor single-valued on the complex plane, and so are not physical operators. In particular, $W(z)$ and $\ol{W}(\zb)$ are ambiguous up to degree $2k$ polynomial shifts in $z$ and $\zb$, respectively,
\begin{align}
\label{eq:Wxamb}
    &W(z)\sim W(z)+c_0+c_1z+\dots +c_{2k}z^{2k},\quad
     \ol{W}(\zb)\sim \ol{W}(\zb)+c_0+c_1\zb+\dots +c_{2k}\zb^{2k}.
\end{align}

The stress tensor components \eqref{eq:Tzzhol}, $T(z)$ and $\ol T(\zb)$ are similarly unphysical before restricting to the physical Hilbert space. Although they are unphysical as local operators, their integrated global charges have a well-defined action on the Hilbert space. Schematically, we write the mode expansions of the stress tensor components as
\begin{subequations}
    \begin{align}
    \label{eq:Tmodeexpansion}
    &T(z)=-\frac{1}{2\pi}\sum_{m=-\infty}^\infty\frac{\ell_m}{z^{m+2}}-\frac{\mathrm{i}}{2\pi}\sum_{m=-\infty}^\infty\frac{b_m}{z^{m+2}}\log z,\\
    &\ol T(\zb)=-\frac{1}{2\pi}\sum_{m=-\infty}^\infty\frac{\tilde{\ell}_m}{\zb^{m+2}}+\frac{\mathrm{i}}{2\pi}\sum_{m=-\infty}^\infty\frac{\tilde{b}_m}{\zb^{m+2}}\log \zb.
\end{align}
\end{subequations}
The logarithmic dependence of these mode expansions characterizes the multi-valuedness of the stress tensor. In particular, the stress tensor components satisfy the monodromy relations,
\begin{subequations}
\begin{align}
\label{eq:monoT}
&T(ze^{2\pi\mathrm{i}})=T(z)+\sum_{|n|\leq k}\( \frac{\mathrm{i}^{k-n}\sqrt{\pi}}{(k-n)!(k+n)!}\frac{p_n}{z^{n-k+1}}\(kz\pp F(z)+(k+1)(k-n)F(z)\)\),\\
\label{eq:monoTb}
    &\ol T(\zb e^{-2\pi\mathrm{i}})=\ol T(\zb)-\sum_{|n|\leq k}\( \frac{\mathrm{i}^{k-n}\sqrt{\pi}}{(k-n)!(k+n)!}\frac{p_{-n}}{\zb^{n-k+1}}\(k\zb\ol\pp \ol F(\zb)+(k+1)(k-n)\ol F(\zb)\)\).
\end{align}
\end{subequations}
The $b_m$ and $\tilde{b}_m$ operators are the coefficients in the Laurent expansions of the second term on the right hand-sides of \eqref{eq:monoT} and \eqref{eq:monoTb}, respectively.

Notably, because the monodromy is proportional to $p_n$, the stress tensor components are single-valued when acting on the physical Hilbert space, i.e. $b_m|\text{phys}\rangle=0$ and $\tilde{b}_m|\text{phys}\rangle=0$ for all $m$. Furthermore, as we show in Appendix \ref{app:b=0}, $b_{-1},b_0,b_1$ and $\tilde{b}_{-1},\tilde{b}_0,\tilde{b}_1$ are identically zero, and so the mode operators $\ell_{-1},\ell_0,\ell_1$ and $\tilde{\ell}_{-1},\tilde{\ell}_0,\tilde{\ell}_1$ are not impacted by the monodromy.

The generators associated with global conformal transformations are,
\begin{subequations}
\begin{align}
    \label{eq:dSglob-1l}
    \ell_{-1}=&\sum_{n=k+1}^{\infty}\sqrt{\frac{(n-k)(n+k+1)}{n(n+1)}}\alpha_{-1-n}\alpha_n+\frac{1}{2}\mathrm{i}^k\frac{(k+2)(2k+1)}{\sqrt{(k+1)(2k+1)!}}p_{k}\alpha_{-k-1}\nonumber\\
    &-\frac{1}{4}\sum_{n=-k}^{k-1}(n-k)p_{-1-n}x_n-\frac{\mathrm{i}}{2}\sum_{n=-k}^{k-1}\frac{(-1)^{k-n}(k+1)}{(k-n)!(k+n)!}p_{-1-n}p_n,\\
    \ell_0=&\sum_{n=k+1}^\infty\alpha_{-n}\alpha_n-\frac{\mathrm{i}}{4}\sum_{|n|\leq k}np_{-n}x_n +\frac{1}{2}\sum_{|n|\leq k}\frac{(-1)^{k-n}(k+1)}{(k-n)!(k+n)!}p_{-n}p_n+a_{\text{dS}},\\
    \ell_{+1}=&\sum_{n=k+1}^{\infty}\sqrt{\frac{(n+k)(n-k-1)}{n(n-1)}}\alpha_{1-n}\alpha_n+\frac{1}{2}\mathrm{i}^{-k}\frac{(k+2)(2k+1)}{\sqrt{(k+1)(2k+1)!}}p_{-k}\alpha_{k+1}\nonumber\\
    &+\frac{1}{4}\sum_{n=-k+1}^{k}(n+k)p_{1-n}x_n+\frac{\mathrm{i}}{2}\sum_{n=-k+1}^{k}\frac{(-1)^{k-n}(k+1)}{(k-n)!(k+n)!}p_{1-n}p_n,
    \end{align}
    \end{subequations}
    for the holomorphic component $T(z)$, and
    \begin{subequations}
        \begin{align}
        \tilde{\ell}_{-1}=&\sum_{n=k+1}^{\infty}\sqrt{\frac{(n-k)(n+k+1)}{n(n+1)}}\tilde{\alpha}_{-1-n}\tilde{\alpha}_n+\frac{1}{2}\mathrm{i}^k\frac{(k+2)(2k+1)}{\sqrt{(k+1)(2k+1)!}}p_{-k}\tilde{\alpha}_{-k-1}\nonumber\\
    &-\frac{1}{4}\sum_{n=-k}^{k-1}(n-k)p_{1+n}x_{-n}-\frac{\mathrm{i}}{2}\sum_{n=-k}^{k-1}\frac{(-1)^{k-n}(k+1)}{(k-n)!(k+n)!}p_{1+n}p_{-n},\\
    \tilde{\ell}_0=&\sum_{n=k+1}^\infty\tilde{\alpha}_{-n}\tilde{\alpha}_n-\frac{\mathrm{i}}{4}\sum_{|n|\leq k}np_{n}x_{-n} +\frac{1}{2}\sum_{|n|\leq k}\frac{(-1)^{k-n}(k+1)}{(k-n)!(k+n)!}p_{-n}p_n+\tilde{a}_{\text{dS}},\\
    \label{eq:dSglob+1r}
    \tilde{\ell}_{+1}=&\sum_{n=k+1}^{\infty}\sqrt{\frac{(n+k)(n-k-1)}{n(n-1)}}\tilde{\alpha}_{1-n}\tilde{\alpha}_n+\frac{1}{2}\mathrm{i}^{-k}\frac{(k+2)(2k+1)}{\sqrt{(k+1)(2k+1)!}}p_{k}\tilde{\alpha}_{k+1}\nonumber\\
    &+\frac{1}{4}\sum_{n=-k+1}^{k}(n+k)p_{-1+n}x_{-n}+\frac{\mathrm{i}}{2}\sum_{n=-k+1}^{k}\frac{(-1)^{k-n}(k+1)}{(k-n)!(k+n)!}p_{-1+n}p_{-n},
\end{align} 
\end{subequations}
for the anti-holomorphic component $\ol{T}(\zb)$. In this Hilbert space we can set to zero all instances of $p_n$, and so these are generators of the Lie algebra of $\mathrm{SL}(2,\mathbb{R})\times\mathrm{SL}(2,\mathbb{R})$, up to normal ordering prescriptions $a_{\text{dS}}$ and $\tilde{a}_{\text{dS}}$,
\begin{subequations}
\begin{align}
    &[\ell_m,\ell_n]=(m-n)(\ell_{m+n}-a_{\text{dS}}\delta_{m+n,0}),\quad m,n=-1,0,1,\\
    &[\tilde{\ell}_m,\tilde{\ell}_n]=(m-n)(\tilde{\ell}_{m+n}-\tilde{a}_{\text{dS}}\delta_{m+n,0}),\quad m,n=-1,0,1,
\end{align}
\end{subequations}
and $[\ell_{m},\tilde{\ell}_n]=0$. These generators act on $W(z)$ and $\ol{W}(\zb)$ as if they are global conformal primaries with holomorphic weights $(h_{W},\ol h_{W})=(-k,0)$ and $(h_{\ol{W}},\ol h_{\ol{W}})=(0,-k)$, up to $x$-mode and logarithmic shifts
\begin{subequations}
\begin{align}
   & [\ell_m,W(z)]\sim z^{m+1}\pp W(z)-k(m+1)z^mW(z),\quad[\tilde{\ell}_m,W(z)]\sim 0,\\
    &[\tilde{\ell}_m,\ol{W}(\zb)]\sim \zb^{m+1}\ol\pp\ol{W}(\zb)-k(m+1)\zb^m\ol{W}(\zb),\quad[\ell_m,\ol{W}(\zb)]\sim 0,
\end{align}
\end{subequations}
where $\sim$ denotes equivalence up to $x$-mode shifts \eqref{eq:Wxamb}, and $p_n$ operator dependence which vanishes in the physical Hilbert space. Let us comment on this transformation law. This is similar to the transformation law of the massless scalar field (corresponding to $k=0$) in two dimensions, which also behaves like a global conformal primary, with holomorphic weights $(h,\ol{h})=(0,0)$, up to constant and logarithmic shifts related to its shift symmetry. The ambiguity for the massless scalar field can be traced back to its logarithmic two-point function, which can be chosen to be
\begin{align}
    \langle\phi(z,\zb)\phi(z',\zb')\rangle=-\frac{1}{4\pi}\log|z-z'|^2,\quad \text{(Massless scalar)},
\end{align}
which does not satisfy global conformal Ward identities consistent with a $(h,\ol{h})=(0,0)$ primary. When $k>0$, $W(z)$ and $\ol{W}(\zb)$ also have logarithmic two-point functions, which can be chosen up to $x$-mode shift ambiguities \eqref{eq:Wxamb} to be,
\begin{subequations}
\begin{align}
    &\langle W(z)W(z')\rangle=(-1)^{k+1}\frac{1}{4\pi}\frac{1}{(2k)!}(z-z')^{2k}\log(z-z'),\\
    &\langle W(z)\ol{W}(\zb ')\rangle=0,\\
    &\langle \ol{W}(\zb)\ol{W}(\zb ')\rangle=(-1)^{k+1}\frac{1}{4\pi}\frac{1}{(2k)!}(\zb-\zb ')^{2k}\log(\zb-\zb').
\end{align}
\end{subequations}
The scalar two-point function can then be obtained from these using \eqref{eq:noncovAnsatz}. For the same reason as the massless scalar field, $W(z)$ and $\ol{W}(\zb)$  are therefore not strictly global conformal primaries with holomorphic weights $(h_{W},\ol h_{W})=(-k,0)$ and $(h_{\ol{W}},\ol h_{\ol{W}})=(0,-k)$. They are instead global conformal primaries up to $x$-mode shift ambiguities.

Because of the transformation properties of the components of $\mathcal{W}^{\mu_1\cdots\mu_k}$ at the quantum level, the scalar field $\phi$ also has the appropriate transformation law \eqref{eq:phiglobconftran}, up to $x$-mode shift ambiguities. 

The components $F(z)$ and $\ol{F}(\zb)$ on the other hand are shift invariant, and behave as global conformal primaries with holomorphic weights $(h_F,\ol{h}_F)=(k+1,0)$ and $(h_{\ol F},\ol h_{\ol F})=(0,k+1)$ in the physical Hilbert space, 
\begin{subequations}
\begin{align}
     &[\ell_m,F(z)]\sim z^{m+1}\pp F(z)+(k+1)(m+1)z^mF(z),\quad[\tilde{\ell}_m,F(z)]\sim 0,\\
    &[\tilde{\ell}_m,\ol F(\zb)]\sim \zb^{m+1}\ol\pp\ol F(\zb)+(k+1)(m+1)\zb^m\ol{F}(\zb),\quad[\ell_m,\ol F(\zb)]\sim 0.
\end{align}
\end{subequations}
To confirm that these generators imply that $F(z)$ and $\ol{F}(\zb)$ admit two-point functions of the form \eqref{eq:FFtwopoint}, the vacuum state must additionally be annihilated by these generators. Indeed, with the vacuum state defined as in \eqref{eq:vacuum}, it is straightforward to see that all global conformal generators annihilate it
\begin{align}
    \ell_m|\Omega\rangle=0,\quad \tilde{\ell}_m|\Omega\rangle=0,\quad m=-1,0,1.
\end{align}
Similarly to $\mathrm{AdS}_2$, the generators $\ell_m$ and $\tilde{\ell}_m$ with $|m|>1$ do not implement the local conformal transformations \eqref{eq:phiconftran}. Additionally in $\mathrm{dS}_2$, they now have terms involving $\alpha_{m-n}x_n$ or $\tilde{\alpha}_{m-n}x_n$ respectively, which can move a state out of the physical Hilbert space. The global conformal generators \eqref{eq:dSglob-1l}--\eqref{eq:dSglob+1r} are therefore the only physically relevant generators constructed from $T(z)$ and $\ol{T}(\zb)$.


\section{Concluding remarks \label{sec: conclusion}}
We have shown that the free scalar field $\phi$ in $\mathrm{dS}_2$ admits a non-standard conformal symmetry when it has the mass squared parameter $m_k^{2}=-k(k+1)R/2$, for each $k\in\mathbb{Z}_{\geq 0}$. The conformal symmetry acts on $\phi$ non-locally, but acts locally in an equivalent description of the theory in terms of a conformal Killing tensor $\mathcal{W}^{\mu_1\cdots\mu_k}$. Because this equivalence is only established after imposing boundary conditions and the equation of motion, the conformal symmetry transformation \eqref{eq:phiglobconftran} is not yet well-defined off-shell. This is related to the fact that the symmetry is only manifest at the level of the equation of motion, and not at the level of the action. 

Having established the conformal symmetry in $\mathrm{dS}_2$ with periodic boundary conditions, it will be interesting to test this in other settings, such as for scalar fields with more general boundary conditions \cite{Higuchi:2022nfy}, or on spacetimes which are locally $\mathrm{dS}_2$. This will test whether the equivalence between a real scalar $\phi$ and $\mathcal{W}^{\mu_1\cdots\mu_k}$ is true beyond $\mathrm{dS}_2$ or $\mathrm{AdS}_2$. This is a new example in a long list of equivalences between field theories in two spacetime dimensions \cite{Coleman:1974bu,Mandelstam:1975hb,Coleman:1975pw,Witten:1984}, and is worth developing further. In a forthcoming work together with collaborators, we find that a similar equivalence holds between a massive Dirac field $\Psi$ with mass squared parameter $m_k^2=-(k+1)^2R/2$ and a spinor with $k+1$ symmetric and traceless tensor indices $\mathcal{Y}^{\mu_1\cdots\mu_{k+1}}$ in $\mathrm{AdS}_2$ \cite{Chen:2027223}.

To make the conformal symmetry manifest at the level of the action, it is necessary to either reconsider how the symmetry acts on the scalar field $\phi$ before the equation of motion is imposed, or start with an alternate formulation. As for the latter choice, one could treat $\mathcal{W}^{\mu_1\cdots\mu_k}$ and $\mathcal{F}_{\mu_1\cdots\mu_{k+1}}$ as independent variables, starting with the action
\begin{align}
    S_{\mathcal{F}\mathcal{W}}=(-1)^{k+1}\int \mathrm{d}^2x\sqrt{-g}\mathcal{F}_{\mu_1\cdots\mu_{k+1}}\nabla^{(\mu_1}\mathcal{W}^{\mu_2\cdots\mu_{k+1})_T}.
\end{align}
Treating both $\mathcal{W}^{\mu_1\cdots\mu_k}$ and $\mathcal{F}_{\mu_1\cdots\mu_{k+1}}$ as Weyl invariant, this action is conformally invariant on any two-dimensional spacetime, with the same traceless stress tensor \eqref{eq:Tmunu}. Furthermore, the equations of motion imply as in the discrete series scalar theory that $\mathcal{W}^{\mu_1\cdots\mu_{k}}$ is a conformal Killing tensor, and $\mathcal{F}_{\mu_1\cdots\mu_{k+1}}$ is covariantly conserved. It is in this setting that one expects the modes of the stress tensor to generate the full Virasoro algebra. This theory would be interesting to study in its own right, and is reminiscent of the $\beta\gamma$ system. The action studied in this paper \eqref{eq:FWaction} may also have connections with boundary reductions of $3d$ higher-spin gravity constructions \cite{Chen:2025xlo}.

As we saw in Section \ref{sec:traceless}, when $\mathcal{W}^{\mu_1\cdots\mu_k}$ and $\mathcal{F}_{\mu_1\cdots\mu_{k+1}}$ are not independent, the stress tensor $T_{\mu\nu}$ \eqref{eq:Tmunu} has peculiar properties. In $\mathrm{dS}_2$, $T(z)$ and $\ol{T}(\zb)$ are not shift invariant, and hence not physical as local operators. They can however be integrated to get well-defined generators of global conformal symmetry transformations, while the rest of the integrated currents $z^{m+1}T(z)$ and $\zb^{m+1}\ol T(\zb)$ with $|m|>1$ do not keep states in the physical Hilbert space. In $\mathrm{AdS}_2$, the generators $\ell_m$ are all well-defined but act non-locally on $\phi(z,\zb)$ for all $|m|>1$, and furthermore do not obey a closed algebra. It could be interesting to determine what set of operators are necessary to define a closed algebra involving all $\ell_m$, and study the resulting algebra.

\section*{Acknowledgments}

The author would like to thank Calvin Y.-R. Chen and Kurt Hinterbichler for feedback on an earlier draft. The research of L.W.L. is supported by the Taiwan NSTC Grant No. 113-2811-M-002-167-MY3 and the Yushan Young Fellowship.

\bibliography{Bibliography}

\appendix
\section{Lorentzian to Euclidean signature for $\mathrm{(A)dS}_2$}
\label{app: LtoE}
In this Appendix, we review the analytic continuations to Euclidean signature used in \cite{Chen:2026kjo} to describe these theories. Together with the Klein--Gordon inner products in Lorentzian signature, the mode expansions and commutation relations reviewed in Section \ref{sec:traceless} can be verified.
\subsection{Anti-de Sitter}
On $\mathrm{AdS}_2$ with radius $L$, its Ricci scalar is $R=-2/L^2$. To describe this spacetime, we write the invariant interval using conformally flat coordinates $(\tau,\rho)$ that covers the global patch,
\begin{align}\label{eq:LorentzAdS}
    \mathrm{d}s^2=-\frac{2}{R}\frac{1}{\cos(\rho)^2}(-\mathrm{d}\tau^2+\mathrm{d}\rho^2),
\end{align}
with $\tau\in\mathbb{R}$ and $\rho\in (-\pi/2,\pi/2)$. This spacetime is topologically equivalent to a strip, with two disjoint conformal time-like boundaries at $\rho=\pm\pi/2$. 

The Klein--Gordon inner product is defined as usual on constant time $\tau$ slices,
\begin{align}
    (\phi_1,\phi_2)_{\mathrm{AdS}_2}\equiv\mathrm{i}\int_{-\pi/2}^{\pi/2} \mathrm{d}\rho\(\ol{\phi_1}\pp_\tau\phi_2-\pp_\tau\ol{\phi_1}\phi_2\).
\end{align}
To go to the complex coordinates \eqref{eq:coords} when $R<0$, we start by analytically continuing $\tau=-\mathrm{i}\tau_\mathrm{E}$ with $\tau_{\mathrm{E}}\in\mathbb{R}$, followed by introducing the complex coordinate
\begin{align}
    z=e^{\tau_{\mathrm{E}}-\mathrm{i}(\rho-\pi/2)},
\end{align}
such that
\begin{align}
    \rho=\frac{\mathrm{i}}{2}\log(z/\zb)+\frac{\pi}{2},\quad \tau_{\mathrm{E}}=\frac{1}{2}\log(z\zb).
\end{align}
This coordinate transformation maps the analytically continued coordinates $(\tau_{\mathrm{E}},\rho)$ to the complex upper half-plane, with the invariant interval
\begin{align}
    \mathrm{d}s^2=\frac{8}{R}\frac{\mathrm{d}z\mathrm{d}\zb}{(z-\zb)^2}.
\end{align}
\subsection{de Sitter}
\label{sapp:dS}
On $\mathrm{dS}_2$ with Hubble scale $H$, its Ricci scalar is $R=2H^2$. To describe this spacetime, we write the invariant interval using conformally flat coordinates $(\eta,\vartheta)$ that covers the global patch,
\begin{align}
     \mathrm{d}s^2=+\frac{2}{R}\frac{1}{\cos(\eta)^2}(-\mathrm{d}\eta^2+\mathrm{d}\vartheta^2),
\end{align}
with $\eta\in(-\pi/2,\pi/2)$ and $\vartheta\in(-\pi,\pi]$. This spacetime is topologically equivalent to a cylinder with two disjoint conformally compactified space-like boundaries at $\eta=\pm\pi/2$.

The Klein--Gordon inner product is defined as usual on constant time $\eta$ slices,
\begin{align}
\label{eq:dSinnerproduct}
    (\phi_1,\phi_2)_{\mathrm{dS}_2}\equiv \mathrm{i}\int_{-\pi}^\pi\mathrm{d}\vartheta\(\ol{\phi_1}\pp_\eta\phi_2-\pp_\eta\ol{\phi_1}\phi_2\).
\end{align}
To go to the complex coordinates \eqref{eq:coords} when $R>0$, we start by analytically continuing $\eta=-\mathrm{i}\eta_\mathrm{E}$. The analytically continued invariant interval is now
\begin{align}
    \mathrm{d}s^2=+\frac{2}{R}\frac{1}{\cosh(\eta_{\mathrm{E}})^2}(\mathrm{d}\eta_{\mathrm{E}}^2+\mathrm{d}\vartheta^2).
\end{align}
We consider the maximal extension of the coordinates with $\eta_{\mathrm{E}}\in\mathbb{R}$, which now describes a sphere $S^2$. Then we introduce the complex coordinate
\begin{align}
    z=e^{\eta_{\mathrm{E}}-\mathrm{i}\vartheta},
\end{align}
such that
\begin{align}
    \vartheta=\frac{\mathrm{i}}{2}\log(z/\zb),\quad \eta_{\mathrm{E}}=\frac{1}{2}\log(z\zb).
\end{align}
This coordinate transformation maps the analytically continued coordinates $(\eta_{\mathrm{E}},\vartheta)$ to the Riemann sphere, with the invariant interval
\begin{align}
    \mathrm{d}s^2=\frac{8}{R}\frac{\mathrm{d}z\mathrm{d}\zb}{(1+z\zb)^2}.
\end{align}
\section{From the Klein--Gordon to conformal Killing tensor equation  \label{app: KGtoCKT}}

In this Appendix, we will prove the intertwining formula \eqref{eq:KGtoCKT} in two-dimensional constant curvature spacetimes when $\mathcal{W}^{\mu_1\cdots\mu_k}$ is symmetric and traceless. This formula is equivalent to the following identity,

\begin{identity}
    On a two-dimensional manifold with constant Ricci scalar $R$, the following identity holds for any symmetric and traceless rank $k$ tensor $\mathcal{W}^{\mu_1\cdots\mu_k}$,
    \begin{align}
        \label{eq:identity1}
        \Delta\nabla_{\mu_1}\cdots\nabla_{\mu_k}\mathcal{W}^{\mu_1\cdots\mu_k} = 2\nabla_{\mu_1}\cdots\nabla_{\mu_{k+1}}\nabla^{(\mu_1}\mathcal{W}^{\mu_2\cdots\mu_{k+1})_T}-\frac{k(k+1)}{2}R\nabla_{\mu_1}\cdots\nabla_{\mu_k}\mathcal{W}^{\mu_1\cdots\mu_k}.
    \end{align}
\end{identity}
\begin{proof}
    First we will commute one of the covariant derivatives from the Laplace--Beltrami operator $\Delta=g^{\mu\nu}\nabla_\mu\nabla_\nu$ through the $k$ covariant derivatives contracting the indices of $\mathcal{W}^{\mu_1\cdots\mu_k}$. For this, we note the two-dimensional commutator identity acting on some rank $p\leq k$ tensor $K^{\lambda_1\cdots\lambda_p}$
    \begin{align}
        [\nabla_\mu,\nabla_\nu]K^{\lambda_1\cdots\lambda_p}=\sum_{i=1}^p\frac{R}{2}\(\delta^{\lambda_i}_{\;\mu}K^{\lambda_1\cdots\;\;\cdots\lambda_p}_{\;\;\;\;\;\;\;\nu}-\delta^{\lambda_i}_{\;\nu}K^{\lambda_1\cdots\;\;\cdots\lambda_p}_{\;\;\;\;\;\;\;\mu}\).
    \end{align}
    Using this successively, we arrive at the expression,
    \begin{align}
        \Delta \nabla_{\mu_1}\cdots\nabla_{\mu_k}\mathcal{W}^{\mu_1\cdots\mu_k} =\nabla_\nu\nabla_{\mu_1}\cdots\nabla_{\mu_k}\nabla^\nu\mathcal{W}^{\mu_1\cdots\mu_k} -\sum_{p=1}^{k}\frac{p}{2}\nabla_{\mu_1}\cdots\nabla_{\mu_p}\(R\nabla_{\mu_{p+1}}\cdots\nabla_{\mu_k}\mathcal{W}^{\mu_1\cdots\mu_k}\).
    \end{align}
    Restricting to manifolds with constant Ricci scalar, this becomes
    \begin{align}
    \label{eq:A4}
        \Delta \nabla_{\mu_1}\cdots\nabla_{\mu_k}\mathcal{W}^{\mu_1\cdots\mu_k} =\nabla_\nu\nabla_{\mu_1}\cdots\nabla_{\mu_k}\nabla^\nu\mathcal{W}^{\mu_1\cdots\mu_k} -\frac{k(k+1)}{4}R\nabla_{\mu_1}\cdots\nabla_{\mu_k}\mathcal{W}^{\mu_1\cdots\mu_k}.
    \end{align}
    Next, we expand the first term on the right hand-side of \eqref{eq:A4} in the explicit complex coordinate system $(z,\zb)$ in \eqref{eq:coords}. In these coordinates, this term expands into four parts
    \begin{align}
        \label{eq:A5}
        \nabla_\nu\nabla_{\mu_1}\cdots\nabla_{\mu_k}\nabla^\nu\mathcal{W}^{\mu_1\cdots\mu_k}=&\,\nabla_z^{k+1}\nabla^z\mathcal{W}^{z\cdots z}+\nabla_{\zb}^{k+1}\nabla^{\zb}\mathcal{W}^{\zb\cdots\zb}\nonumber\\
        &\,+\nabla_{\zb}\nabla_{z}^{k}\nabla^{\zb}\mathcal{W}^{z\cdots z}+\nabla_z\nabla_{\zb}^k\nabla^z\mathcal{W}^{\zb\cdots\zb}
    \end{align}
    We recognize the first line on the right hand-side as $\nabla_{\mu_1}\cdots\nabla_{\mu_{k+1}}\nabla^{(\mu_1}\mathcal{W}^{\mu_2\cdots\mu_{k+1})_T}$ in coordinate form. The second line requires more care. Using $\nabla^{z}=g^{z\zb}\nabla_{\zb}$ and $\nabla^{\zb}=g^{\zb z}\nabla_z$,
    \begin{align}
        \label{eq:A6}\nabla_{\zb}\nabla_{z}^{k}\nabla^{\zb}\mathcal{W}^{z\cdots z}+\nabla_z\nabla_{\zb}^k\nabla^z\mathcal{W}^{\zb\cdots\zb}=&\,g^{\zb z}\nabla_{\zb}\nabla_z\nabla_z^k\mathcal{W}^{z\cdots z}+g^{z\zb}\nabla_z\nabla_{\zb}\nabla_{\zb}^k\mathcal{W}^{\zb\cdots\zb}\nonumber\\
        =&\,\frac{1}{2}\Delta\nabla_{\mu_1}\cdots\nabla_{\mu_k}\mathcal{W}^{\mu_1\cdots\mu_k},
    \end{align}
    where we have used the fact that $g^{z\zb}\nabla_z\nabla_{\zb}=g^{\zb z}\nabla_{\zb}\nabla_{z}=g^{z\zb}\pp\ol\pp=\frac{1}{2}\Delta$ when acting on a rank 0 holomorphic tensor. Combining \eqref{eq:A5} and \eqref{eq:A6} with \eqref{eq:A4}, we get \eqref{eq:identity1}.
\end{proof}
\section{Monodromy of the stress tensor in $\mathrm{dS}_2$}
\label{app:b=0}
As described in Section \ref{ssec:dS2}, the stress tensor components $T(z)$ and $\ol{T}(\zb)$ in $\mathrm{dS}_2$ are not single-valued before restricting to the physical Hilbert space. In this Appendix, we characterize the monodromy \eqref{eq:monoT} and \eqref{eq:monoTb} further by giving an explicit expression for $b_m$ and $\tilde{b}_m$, and in particular verify that $b_{-1},b_0,b_1=0$. Another calculation can be done to show that $\tilde{b}_{-1},\tilde{b}_0,\tilde{b}_1=0$, but because of how similar $b_m$ and $\tilde{b}_m$ are, we omit it.

Writing out the mode expansion for $F(z)$ as in \eqref{eq:Fmode}, we have
\begin{align}
    kz\pp F(z)+(k+1)(k-n)F(z)=-\frac{\mathrm{i}}{\sqrt{4\pi}}\sum_{m=-\infty}^{\infty}(n+k(m+n))\frac{F_m}{z^{m+k+1}}.
\end{align}
Using this we can write out the Laurent expansion of the operator appearing in the monodromy in \eqref{eq:monoT}
\begin{align}
    \sum_{|n|\leq k}\bigg( \frac{\mathrm{i}^{k-n}\sqrt{\pi}}{(k-n)!(k+n)!}\frac{p_n}{z^{n-k+1}}&\(kz\pp F(z)+(k+1)(k-n)F(z)\)\bigg)=\nonumber\\
   & =-\frac{\mathrm{i}}{2}\sum_{m=-\infty}^{\infty}\sum_{|n|\leq k}\mathrm{i}^{k-n}\frac{n+km}{(k-n)!(k+n)!}\frac{F_{m-n}p_n}{z^{m+2}}.
\end{align}
Matching this with the definition of $b_m$ in \eqref{eq:Tmodeexpansion}, we find
\begin{align}
\label{eq:explbm}
    b_m=-\frac{\mathrm{i}}{2}\sum_{|n|\leq k}\mathrm{i}^{k-n}\frac{(n+km)}{(k-n)!(k+n)!}F_{m-n}p_n.
\end{align}
Similarly, $\tilde{b}_m$ can be written out in terms of the corresponding Laurent expansion of $\ol{F}(\zb)$ with mode operators $\tilde{F}_m$
\begin{align}
\label{eq:expltbm}
    \tilde{b}_m=-\frac{\mathrm{i}}{2}\sum_{|n|\leq k}\mathrm{i}^{k-n}\frac{(n+km)}{(k-n)!(k+n)!}\tilde{F}_{m-n}p_{-n}.
\end{align}

Restricting to $m=-1,0,1$, the only modes from $F(z)$ which appear in \eqref{eq:explbm} are $F_n=-\mathrm{i}^{k+n}p_n$ with $n\leq k$. Indeed, the operators $F_{k+1}=-\mathrm{i}^k\sqrt{\frac{(2k+1)!}{k+1}}\alpha_{k+1}$ and $F_{-k-1}=-\mathrm{i}^k\sqrt{\frac{(2k+1)!}{k+1}}\alpha_{-k-1}$ that could have appeared in $b_m$ when $m=1$ and $m=-1$ respectively do not contribute because of the factor $(n+km)$ in the summand.

Plugging $F_{m-n}=-\mathrm{i}^{k+m-n}p_{m-n}$ into \eqref{eq:explbm} for $m=-1,0,1$, it is easy to verify that $b_{-1},b_0,b_1=0$.

\end{document}